\documentclass[12pt,journal,draftclsnofoot,onecolumn,letterpaper]{IEEEtran} 
\usepackage{amsthm}
\usepackage{amssymb}
\usepackage{amsmath}
\usepackage{bm}
\usepackage{amsfonts}
\usepackage{graphicx}
\usepackage{subfigure}
\usepackage{cite}
\usepackage[numbers]{natbib}
\usepackage{enumerate}
\usepackage{url}
\usepackage[ruled]{algorithm2e}
\usepackage{mathtools}
\usepackage{epstopdf}
\usepackage{dsfont}
\usepackage{float}

\allowdisplaybreaks[4]
\begin{document}

\newtheorem{definition}{\bf ~~Definition}
\newtheorem{observation}{\bf ~~Observation}
\newtheorem{theorem}{\bf ~~Theorem}
\newtheorem{proposition}{\bf ~~Proposition}
\newtheorem{remark}{\bf ~~Remark}

\title{Integrated Satellite-HAP-Terrestrial Networks for Dual-Band Connectivity\vspace{-0.1cm}}
\author{
\IEEEauthorblockN{
\normalsize{Wenwei Zhang},
\normalsize{Ruoqi Deng},
\normalsize{Boya Di},
\normalsize{and Lingyang Song}\\}
\IEEEauthorblockA{Department of Electronics, Peking University, Beijing, China.}
\\
\vspace{-1.0cm}
}
\maketitle

\begin{abstract}
The recent development of high-altitude platforms (HAPs) has attracted increasing attention since they can serve as a promising communication method to assist satellite-terrestrial networks. In this paper, we consider an integrated three-layer satellite-HAP-terrestrial network where the HAP support dual-band connectivity. Specifically, the HAP can not only communicate with terrestrial users over C-band directly, but also provide backhaul services to terrestrial user terminals over Ka-band. We formulate a sum-rate maximization problem and then propose a fractional programming based algorithm to solve the problem by optimizing the bandwidth and power allocation iteratively. The closed-form optimal solutions for bandwidth allocation and power allocation in each iteration are also derived. Simulation results show the capacity enhancement brought by the dual-band connectivity of the HAP. The influence of the power of the HAP and the power of the satellite is also discussed.
\end{abstract}


\newpage

\section{Introduction}
The explosion of data traffic and users' demand for reliable and high-capacity connectivities pose great challenges to traditional wireless communication systems. The existing terrestrial communication system with unsolved issues such as the short of available bandwidth and the limited backhaul capacity of terrestrial small cells can hardly meet such transmission demand \cite{b01}, . To make up for such deficiency, low-earth-orbit (LEO) satellite networks operating over high-frequency bands have attracted increasing attention since they can provide wide-bandwidth and high-capacity data \cite{b0}. However, severe path loss and long communication delay remain to be the development bottleneck of satellite networks.\\
\indent Fortunately, the emergence of the high-altitude platform (HAP) overcomes such deficiencies of satellite communications. HAPs are stations located on aerial vehicles at an altitude of 20 to 50 km, aiming at exploiting the potential benefits of intermediate altitudes between terrestrial networks and satellite networks. The integration of HAPs and terrestrial networks can provide high-capacity data services, with a reduction in delay and complexity compared with satellite networks \cite{b1}. Therefore, HAPs can serve as a promising communication method to assist satellite-terrestrial networks \cite{b2}. Specifically, HAPs support dual-band connectivity. \emph{First}, it can communicate with terrestrial users over C-band directly. \emph{Second}, it can provide ground user terminals (UTs) with data backhaul services over Ka-band \cite{b3}. The UT acts as an access point that can transmit users' data to the core network via LEO-based backhaul or HAP-based backhaul over Ka-band \cite{b31}.\\
\indent In the literature, various aspects have been considered for HAP-based networks, such as coverage optimization \cite{b4}, resource allocation \cite{b5}, and capacity analysis \cite{b6}. In \cite{b4}, a HAP coverage optimization algorithm has been developed for coverage maximization. In~\cite{b5}, a resource allocation scheme has been proposed for HAP-aided OFDMA multicast systems. In \cite{b6}, a three-layer satellite-HAP-terrestrial architecture model has been proposed for capacity improvement in a HAP network. However, most existing works only focus on the two-layer HAP-terrestrial networks, and only a few consider the three-layer satellite-HAP-terrestrial networks. Nevertheless, the works about the three-layer networks do not consider the dual-band connectivity of the HAP. \\
\indent Different from the above works, in this paper, we consider an integrated three-layer satellite-HAP-terrestrial network. Each terrestrial user can access the network via a HAP or a UT. The HAP can connect to the core network over C-band directly, while the UT connects the core network via LEO-based backhaul or HAP-based backhaul over Ka-band. The network aims to maximize the sum rate of all users. Therefore, the backhaul selection and the bandwidth and power allocation should be jointly optimized subject to the backhaul capacity constraint of each UT as well as that of the HAP.\\
\indent New challenges have arisen in such a three-layer network. On the one hand, due to the constraints of the dynamic backhaul capacity bandwidth allocation over C-band and backhaul selection over Ka-band are coupled. On the other hand, since the two methods for users to connect with the core network through HAP over C-band and Ka-band share the total power of HAP, there is a tradeoff between the sum rate and the backhaul capacity of the HAP. To tackle these challenges, we propose an iterative sum-rate maximization algorithm to solve the problem. The main contribution of this paper can be summarized as follows. (1) We consider an integrated three-layer satellite-HAP-terrestrial network where the HAP serves in two different methods. (2) We formulate a sum-rate maximization problem and then design a joint optimization algorithm to solve the problem. (3) Simulations results show the capacity enhancement brought by the dual-band connectivity of the HAP. The influence of the power of the HAP and the power of the satellite is also evaluated.\\
\indent The rest of this paper is organized as follows. In Section \ref{SM}, we describe the model of a satellite-HAP-terrestrial network. In Section \ref{PFA}, we formulate a sum-rate maximization problem and decompose it into two subproblems. A joint optimization problem is then proposed to solve the two subproblems iteratively in Section \ref{JS}. The simulation results are presented in Section \ref{SR}. Finally, the conclusion is drawn in Section \ref{Con}.
\section{System Model}\label{SM}
In this section, we first introduce an integrated satellite-HAP-terrestrial network. Then we describe the models of the HAP-user communications, HAP-based backhaul and LEO-based backhaul, respectively.
\vspace{-1pt}
\subsection{Scenario Description}
\begin{figure}[!htbp]
\centering
\includegraphics[width=5.4in]{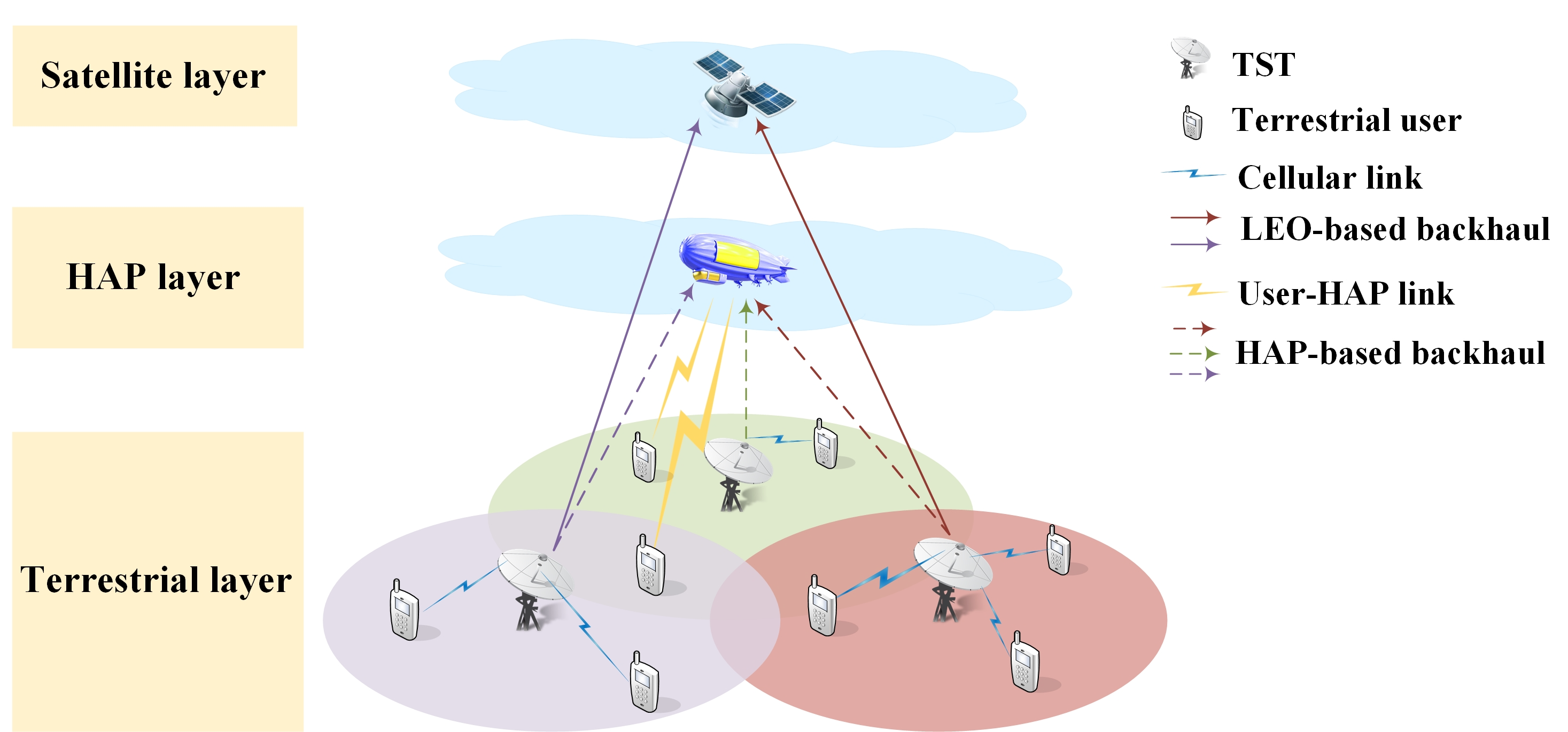}
\caption{System model of the integrated satellite-HAP-terrestrial network}
\label{Fig}
\end{figure}

Consider a downlink integrated network as shown in Fig. \ref{Fig}, where there are one LEO satellite, one HAP, and $M$ UTs each serving a group of terrestrial users. Both LEO satellite and HAP can provide Ka-band backhaul connections to terrestrial users. For the downlink transmission, each user can receive data from the core network via the following three ways: 1) The HAP directly transmits data from the core network to terrestrial users over C-band. 2) The HAP first transmits data from the core network to the UT via the HAP-based backhaul link over Ka-band. The UT then forwards the received data to terrestrial users over C-band. 3) The satellite first transmits data from the core network to the UT via the satellite-based backhaul link over Ka-band. The UT then forwards the received data to terrestrial users over C-band.
\\
\vspace{-1pt}
\subsection{Transmission Model for HAP-User Communications}
As for HAP-user communications, the data is transmitted directly from the HAP to terrestrial users over C-band. Without loss of generality, we assume that the HAP and the UTs share the same C-band frequency resource pool, which can be divided into $J$ orthogonal channels. Both HAP and UTs allocate orthogonal frequency resources to terrestrial users. Denote the set of UTs that as $\mathcal{M}=\{1,2,\cdots ,m,\cdots ,M\}$, the set of users that UT $m$ serves as $\mathcal{N}=\{1,2,\cdots ,n,\cdots ,N_m\}$, the power of each UT as $P_{T}$, and the power allocated by UT $m$ to its user $n$ as $P_{m,n}^{T}$ , which satisfies
\begin{equation}
\sum\limits_{n=1}^{N_m}P_{m,n}^{T}=P_{T}.
\end{equation}
Therefore, the interference in the HAP-user link over C-band comes from the UT-user links sharing the same C-band spectrum. The interference received by HAP-user $n$ of UT $m$ link can then be given by
\begin{equation}
I_{m,n}^{H}=\sum\limits_{\substack{n^{\prime}\in N_m,\\n^{\prime}\neq n}}\frac 1JP_{T,m,n^{\prime}}\vert h_{m,n,T_m}\vert^2+\sum\limits_{m^{\prime}\neq m}\sum\limits_{n^{\prime}=1}^{N_{m^{\prime}}}\vert h_{m,n,T_{m^{\prime}}}\vert^2
\end{equation}
where $h_{m,n,T_m}$ is the channel coefficient between user $n$ of UT $m$ and UT $m$. Considering the large-scale fading, $h_{m,n,T_{m}}$ can be given by $\vert h_{m,n,T_m}\vert^2=d_{m,n,T_m}^{-\varepsilon}$, where $d_{m,n,T_m}$ is the distance between UT $m$ and its user $n$, and $\varepsilon$ represents the path loss exponent.\\
\indent Denote the total power of the HAP as $P_H$. Since the HAP works over both C-band spectrum and Ka-band spectrum, the power allocated to user $n$ of UT $m$ for C-band communication is denoted as $P_{m,n}^{H}$, and the power allocated to each Ka-band backhaul is denoted as $P_{H, Ka}$,
which satisfies
\begin{equation}
\sum\limits_{m=1}^{M}\sum\limits_{n=1}^{N_m}P_{m,n}^{H}+P_{H,Ka}\cdot M=P_H.
\end{equation}
Therefore, the achievable throughput of terrestrial user $n$ of UT $m$ served by HAP can then be given by
\begin{align}
R_{m,n}^{H}=\frac{B_{H}}{\sum\limits_{m=1}^{M}N_m}\log_2\left(1+\frac{P_{m,n}^{H}\vert h_{m,n,H}\vert^2}{\sigma_{c}^2+I_{m,n}^{H}}\right),\label{RH}
\end{align}
where $B_{H}$ is the C-band spectrum that is assigned to the HAP, $h_{m,n,H}$ is the channel coefficient
between user $n$ of UT $m$ and HAP, $\sigma_c^2$ is the additive white Gaussian noise (AWGN) variance, and $I_{m,n}^{H}$ is the interference suffered by HAP-user $n$ of UT $m$ link. Therefore, the total throughput of terrestrial users can be given by $R_H=\sum\limits_{m=1}^{M}\sum\limits_{n=1}^{N_m}R_{m,n}^{H}$.

\subsection{Transmission Model for UT-User Communications}
As for UT-user communications, the HAP or the satellite first transmits data from the core network to the UT via the backhaul link over Ka-band. The UT then forwards the received data to terrestrial users over C-band. The satellite and HAP allocate orthogonal frequency resources to the UT within its coverage for the backhaul link over Ka-band. Similarly, the interference in the UT $m$-terrestrial user $n$ of UT $m$ link comes from the HAP-user link and other UT-user links using the same frequency resource, which can be given by
\begin{equation}
I_{m,n}^{T}=\sum_{m^{\prime}=1}^{M}\sum_{n^{\prime}=1}^{N_{m^{\prime}}}\frac 1JP_{m^{\prime},n^{\prime}}^{H}\vert h_{m,n,H}\vert^2\notag+\sum_{m^{\prime}=m}\sum_{n^{\prime}=1}^{N_{m^{\prime}}}\frac 1JP_{m^{\prime},n^{\prime}}^{T}\vert h_{m,n,T_{m^{\prime}}}\vert^2.
\end{equation}
\indent The achievable throughput of terrestrial user $n$ of UT $m$
$R_{m,n}^{T}$ is then given by
\begin{align}
R_{m,n}^{T}=\frac{B_{m}^{T}}{N_m}\log_2\left(1+\frac{P_{m,n}^{T}\vert h_{m,n,T}\vert^2}{\sigma_c^2+I_{m,n}^{T}}\right)\label{RT},
\end{align}
where $B_{m}^{T}$ is the bandwidth for UT $m$-user communications over C-band, $I_{m,n}^{T}$ is the
interference suffered by UT $m$-user $n$ of UT $m$ link, $h_{m,n,T}$ is the channel coefficient between
user $n$ of UT $m$ and UT $m$, and $\sigma_c^2$ is the additive white Gaussian noise (AWGN) variance at
each user. Therefore, the total throughput of terrestrial users $R_T$ is $R_T=\sum_{m=1}^{M}\sum_{n=1}^{N_m} R_{m,n}^{T}$.
\subsubsection{HAP-based backhaul}
The backhaul capacity of HAP-based backhaul over Ka-band can be given by
\begin{equation}
C_{TH}=\sum\limits_{m=1}^{M_H}\frac{B_{Ka}}{M}\log_2\left(1+\frac{P_{H,Ka}G_{T_{m},H}\vert h_{T_{m},H}\vert^2}{\sigma^2} \right)\label{CTH},
\end{equation}
where $M_{H}$ is the number of UT linked to the HAP, $G_{T_m,H}$ is the antenna
gain between UT $m$ and HAP, $\sigma^2$ is the AWGN variance at
the HAP, $P_{H,Ka}$ is the power of HAP allocated to each Ka-band backhaul. The channel coefficient between UT $m$ and HAP $\vert h_{T_m,H}\vert^2 = d_{T_m,H}^{-\varepsilon}$, where $d_{T_m,H}$ is the distance between UT $m$ and HAP, and $\varepsilon$ represents the path loss exponent.
\subsubsection{LEO-based backhual}
The backhaul capacity of LEO-based backhaul $C_{TS}$ can be given by
\begin{equation}
C_{TS}=\sum\limits_{m=1}^{M_S}\frac{B_{Ka}}{M}\log_2\left(1+\frac{P_{SAT}G_{T_m,S}\vert h_{T_m,S}\vert^2}{\sigma_{s}^2\cdot M_S}\right),\label{CTS}
\end{equation}
where $M_{S}$ is the number of UT linked to the satellite, $P_{SAT}$ is the power of satellite and $\sigma_{s}^2$ is the additive white Gaussian noise (AWGN) variance. $G_{T_m,S}$ is the antenna gain between UT $m$ and satellite, $\vert h_{T_m,S}\vert^2=d_{T_m,S}^{-\varepsilon}$ is the channel coefficient between satellite and UT $m$, where $d_{T_m,S},$ is the distance between UT $m$ and satellite, and $\varepsilon$ represents the path loss exponent.
\section{Promblem Formulation and Decomposition}\label{PFA}
In this section, we formulate the sum-rate maximization problem and decompose it into two subproblems. The overall algorithm framework to solve the problem is then developed.
\subsection{Promblem Formulation}
The sum-rate maximization problem can be formulated as
\begin{align}
\max\limits_{\substack{\{ B_{m}^{T}, B_{H}, P_{m,n}^{H},\\P_{Ka}^{H}, P_{m,n}^{T}, M_H, M_S \}}}&\sum\limits_{m=1}^{M}\sum\limits_{n=1}^{N_{m}}R_{m,n}^{H}+R_{m,n}^{T},\label{pro1}\\
s.t.\qquad\quad&(\ref{CTH}),(\ref{CTS}),\notag\\&R_T\leq C_{TH}+C_{TS},\quad R_{H}\leq C_H, \tag{\ref{pro1}{a}}\label{pro1a}\\
&\sum\limits_{n=1}^{N_{m}}P_{m,n}^{T} = P_{T},\tag{\ref{pro1}{b}}\label{pro1c}\\
&\sum\limits_{m=1}^{M}\!\sum\limits_{n=1}^{N_{m}}\!P_{m,n}^{H}\!+\!P_{H,Ka}\!\cdot\!M_{H}=P_{H},\tag{\ref{pro1}{c}}\label{pro1d}\\
&\sum\limits_{m=1}^{M}B_{m}^{T}+B_{H}=B_C.\tag{\ref{pro1}{d}}\label{pro1e}
\end{align}
Constraint (\ref{pro1a}) implies that the users’ total throughput in UT-user link $R_T$ should be no larger than the sum of LEO-based backhaul capacity $C_{TS}$ and the HAP-based backhaul capacity $C_{TH}$. The users’ total throughput in the HAP-user link $R_H$ should also be no larger than the backhaul capacity of HAP over C-band $C_H$. Constraints (\ref{pro1c}) and (\ref{pro1d}) imply transmit power allocation of the UT and HAP, respectively.
$B_C$ is the C-band assigned to the whole communication system. The orthogonal use of the C-band spectrum for the HAP and the UT is guaranteed by constraint (\ref{pro1e}).
\subsection{Promblem Decomposition}
\indent To maximize the sum rate, the backhaul capacity over Ka-band also needs to be optimized. Note that the sum-rate maximization problem and the backhaul capacity maximization problem are coupled through the power that the HAP allocates to Ka-band, i.e., $P_{H,Ka}$. To decouple these two problems, we first optimize $P_{H,Ka}$ by line search method, such that for each fixed $P_{H,Ka}$, the initial problem can be decomposed into two subproblems, i.e., \textit{(p1)} backhaul selection problem; \textit{(p2)} bandwidth and power allocation problem, as given below:\vspace{2mm}\\
\emph{(p1) Backhaul Selection:}
\begin{align}
\max\limits_{\{M_H, M_S\}}C_{TH}+C_{TS},\quad s.t.\quad P_{H,Ka}\geq 0. \label{e0}
\end{align}
\noindent\emph{(p2) Bandwidth and Power Allocation:}
\begin{align}
\max\limits_{\substack{\{ B_{m}^{T}, B_{H}, P_{m,n}^{H},P_{m,n}^{T}\}}}& \sum\limits_{m=1}^{M}\sum\limits_{n=1}^{N_{m}}R_{m,n}^{H}+R_{m,n}^{T},\label{pro2}\\
s.t.\qquad\quad &(\ref{pro1a})\!-\!(\ref{pro1e}),\notag\\
&R_{T}\leq \max\{C_{TH}+C_{TS}\}\tag{\ref{pro2}a}.\label{pro2a}
\end{align}
\indent The overall sum-rate maximization algorithm is summarized in Algorithm \ref{alg1}, where the problem (p1) and problem (p2) are optimized iteratively. Algorithms \ref{alg2} and \ref{alg3} mentioned in Algorithm \ref{alg1} will be elaborated in the following section.
\begin{algorithm}[!htbp]
	\label{alg1}
	\caption{Overall Iterative Algorithm Framework}
	\KwIn{Total power of HAP $P_{H}$;}
	\textbf{Initialization}: $P_{H,Ka} = 0$;\\
	\Repeat{Search all $P_{H,Ka}$}{
	\textit{Step 1.} For a fixed $P_{H,Ka}$, compute the maximum backhaul capacity by Algorithm \ref{alg2};\\
	\textit{Step 2.} Solve the problem (p2) by Algorithm \ref{alg3} based on the maximum backhaul capacity given in \textit{Step 1} and compute the sum rate;\\
	\textit{Step 3.} Update the maximum sum rate and $P_{H,Ka}$;
	}	
	\KwOut{The maximum sum rate $R_{sum}$, $B_{m}^{T}$, $B_{H}$, $P_{m,n}^{H}$, $P_{m,n}^{T}$, $M_H$, $M_S$;}
\end{algorithm}
\section{Backhaul Selection, Bandwidth and Power Allocation Optimization Algorithms}\label{JS}
In this section, we present the backhaul selection optimization algorithm, the bandwidth and power allocation algorithm separately.
\subsection{Backhaul Selection Optimization}\label{Backhaul}
Note that the value of $C_{TH}+C_{TS}$ is determined by whether each UT chooses the satellite or the HAP for Ka-band backhaul based on (\ref{CTH}) and (\ref{CTS}). To solve (p1), we utilize dynamic programming to maximize the backhaul capacity. Specifically, for initialization, we assume that all the UTs are connected with the satellite. For each possible value of $M_S=m_s$, through sorting the increment of backhaul capacity of each UT, denoted by $\Delta C_m$, when it connects with the HAP instead of the satellite, i.e.,\\
\begin{align}
\Delta C_m = \frac{B_{Ka}}{M}\log_2\left(\frac{1+\frac{P_{H,Ka}G_{T_{m},H}\vert h_{T_{m},H}\vert^2}{\sigma^2}}{1+\frac{P_{SAT}G_{T_m,S}\vert h_{T_m,S}\vert^2}{\sigma_{s}^2\cdot m_s}}\right).\label{dc}
\end{align}\\
From high to low, we can choose the first $m_s$ UTs to connect with HAP, such that the value of $C_{TH}+C_{TS}$ is maximized at $M_S=m_s$. Finally, by searching all possible values of $M_S$, we can obtain the maximum backhaul capacity over Ka-band. The whole backhaul capacity maximization algorithm is summarized in Algorithm \ref{alg2}.
\begin{algorithm}[!htbp]
	\label{alg2}
	\caption{Backhaul Selection}
	\KwIn{Power allocated to each Ka-band backhual $P_{H,Ka}$;}
	\While{$m_s \leq M_S$}{
	\textbf{(1)} For each UT $m$, compute $\Delta C_m$ by (\ref{dc});\\
	\textbf{(2)} Sort UTs by $\Delta C_m$ from high to low and choose the first $m_s$ UTs to connect with HAP;\\
	\textbf{(3)} Compute $C_{TH}+C_{TS}$ by (\ref{CTH}) and (\ref{CTS}) and update the $\max\{C_{TH} +C_{TS}\}$;\\
	\textbf{(4)} $m_s = m_s + 1$;
	}
    \KwOut{$M_H$, $M_S$, $\max\{C_{TH} +C_{TS}\}$;}
\end{algorithm}

\subsection{Bandwidth and Power Allocation Optimization}\label{Reformulations}
To solve (p2), we introduce several auxiliary variables to recast the $R_{m,n}^{H}$ and $R_{m,n}^{T}$, such that the non-convex problem (p2) can be converted into the convex optimization problem, which can be given by
\begin{align}
\max\limits_{\{\bm{x}, \bm{\gamma}, \bm{y} \}} \quad &\sum\limits_{m=1}^{M}\sum\limits_{n=1}^{N_{m}}Q_{m,n}^{H}+Q_{m,n}^{T},\label{cop}\\
s.t.\qquad&(\ref{pro1a})-(\ref{pro1e})(\ref{pro2a}),\notag
\end{align}
where
\begin{equation}\label{QHmn}
\begin{split}
Q_{m,n}^H\!&=\!\frac{B_{H}\left(\log_2\left(1\!+\!\gamma_{m,n}^{1}\right)\!-\!\gamma_{m,n}^{1}\right)}{\sum_m N_m}\!+\!2y_{m,n}^{1}\sqrt{\frac{B_{H} P_{m,n}^{H}\!\left|h_{m,n,H}\right|^2\!\left(1\!+\!\gamma_{m,n}^{1}\right)}{\sum_m N_m}}\\
&\quad \!-\!{y_{m,n}^{1}}^2\left(P_{m,n}^{H}\left|h_{m,n,H}\right|^2\!+\!\sigma_c^2\!+\!I_{m,n}^{H}\right),
\end{split}
\end{equation}
and
\begin{equation}\label{QTmn}
\begin{split}
Q_{m,n}^{T}\!&=\!\frac{B_{m}^{T}\left(\log_2\left(1\!+\!\gamma_{m,n}^{2}\right)\!-\!\gamma_{m,n}^{2}\right)}{N_m}\!+\!2y_{m,n}^{2}\sqrt{\frac{B_{m}^{T} P_{m,n}^{T}\!\left|h_{m,n,T_m}\right|^2\!\left(1\!+\!\gamma_{m,n}^{2}\right)}{N_m}}\\
	&\quad \!-\!{y_{m,n}^{2}}^{2}\left(P_{m,n}^{T}\left|h_{m,n,T_m}\right|^2\!+\!\sigma_c^2\!+\!I_{m,n}^{T}\right).
\end{split}
\end{equation}

\noindent $\bm{x}$ include variables to be optimized, i.e., $B_{m}^{T}$, $B_{H}$, $P_{m,n}^{H}$, $P_{m,n}^{T}$. $\gamma_{m,n}^{1}$, $\gamma_{m,n}^{2}$ and $y_{m,n}^{1}$, $y_{m,n}^{2}$ are the auxiliary variables corresponding to each $(m,n)$ pair. A constructive proof can be found in \cite{b9}.\\
\indent To solve the reformulated problem (\ref{cop}), we optimize the variables $\bm{x}$, $\gamma_{m,n}^{1}$, $\gamma_{m,n}^{2}$, $y_{m,n}^{1}$, $y_{m,n}^{2}$ in an iterative manner. The detailed sum-rate optimization scheme is given as below.\vspace{2mm}
\subsubsection{Optimization of Auxiliary Variables}
When all the other variables are fixed, By setting $\frac{\partial Q_{m,n}^{H}(\bm{x},\gamma_{m,n}^{1},y_{m,n}^{1})}{\partial y_{m,n}^{1}}=0$ and $\frac{\partial Q_{m,n}^{T}(\bm{x},\gamma_{m,n}^{2},y_{m,n}^{2})}{\partial y_{m,n}^{2}}=0$, the optimal $y_{m,n}^{1}$ and $y_{m,n}^{2}$ have an explicit solution given in (\ref{y1star}) and (\ref{y2star}), respectively.\vspace{2mm}\\
\begin{subequations}
\label{y_star}
\begin{equation}
{y_{m,n}^{1}}^{\ast}\!=\!\frac{\sqrt{\frac{B_{H}}{\sum_{m=1}^M N_m}P_{m,n}^{H}\left|h_{m,n,H}\right|^2\left(1\!+\!\gamma_{m,n}^{1}\right)}}{P_{m,n}^{H}\left|h_{m,n,H}\right|^2+\sigma_c^2+I_{m,n}^{H}},\label{y1star}
\end{equation}
\begin{equation}
{y_{m,n}^{2}}^{\ast}\!=\!\frac{\sqrt{\frac{B_{m}^{T}}{N_m}P_{m,n}^{T}\left|h_{m,n,T_m}\right|^2\left(1+\gamma_{m,n}^{2}\right)}}{P_{m,n}^{T}\left|h_{m,n,T_m}\right|^2\!+\!\sigma_c^2\!+\!I_{m,n}^{T}}.\label{y2star}
\end{equation}
\end{subequations}
\vspace{2mm}\\
After substituting the ${y_{m,n}^{1}}^{\ast}$ and ${y_{m,n}^{2}}^{\ast}$ in (\ref{QHmn}) and (\ref{QTmn}), we can obtain the optimal ${\gamma_{m,n}^{1}}^{\ast}$ and ${\gamma_{m,n}^{2}}^{\ast}$ by setting $\frac{\partial Q_{m,n}^{H}(\bm{x},\gamma_{m,n}^{1})}{\partial \gamma_{m,n}^{1}}=0$ and $\frac{\partial Q_{m,n}^{T}(\bm{x},\gamma_{m,n}^{2})}{\partial \gamma_{m,n}^{2}}=0$ as \vspace{2mm}
\begin{equation}
{\gamma_{m,n}^{1}}^{\ast} = \frac{P_{m,n}^{H}\vert h_{m,n,H}\vert^2}{\sigma_{c}^2+I_{m,n}^{H}},\quad
{\gamma_{m,n}^{2}}^{\ast} = \frac{P_{m,n}^{T}\vert h_{m,n,T}\vert^2}{\sigma_c^2+I_{m,n}^{T}}.\label{gamma_star}
\end{equation}
\vspace{2mm}

\subsubsection{Optimization of Original Variables}
\indent We utilize Lagrange multiplier to relax constrains (\ref{pro1c}), (\ref{pro1d}) and (\ref{pro1e}). The Lagrange function can be expressed as
\begin{equation}\label{Lan}
\begin{split}
L(\bm{x},\bm{\gamma},\bm{y},\lambda_B,\lambda_{Tm},\lambda_{H})=&\sum\limits_{m=1}^{M}\sum\limits_{n=1}^{N_{m}}\left(Q_{m,n}^{H}+Q_{m,n}^{T}\right)+\lambda_B(B_C-B_{H}-\sum_{n=1}^M B_{m}^{T})\\
&+\sum_{m=1}^{M}\!\lambda_{T,m}(P_{T}\!-\!\sum_{n=1}^{N_m}P_{m,n}^{T})
\!+\!\lambda_H(P_H\!-\!P_{H,Ka}M\!-\!\sum_{m=1}^M\sum_{n=1}^{N_m}P_{m,n}^{H}),
\end{split}
\end{equation}
where $\lambda_{B}$, $\lambda_{T,m}$, and $\lambda_{H}$ are Lagrange multipliers associated with constraints (\ref{pro1e}), (\ref{pro1c}) and (\ref{pro1d}), respectively.
Setting $\frac{\partial L}{\partial B_{H}}=0$, we have
\begin{align}
&-\lambda_B+X_{B_{H}}+Y_{B_{H}}\times\frac{1}{\sqrt{B_{H}}}=0,\label{1a}
\end{align}
where
\begin{align}
&X_{B_{H}}=\sum\limits_{m=1}^M\sum\limits_{n=1}^{N_m} \frac{\log_2(1+\gamma_{m,n}^{1})-\gamma_{m,n}^{1}}{\sum N_m},\\
&Y_{B_{H}}\!=\!\sum\limits_{m=1}^M\!\sum\limits_{n=1}^{N_m} y_{m,n}^{1}\sqrt{\frac{P_{m,n}^{H}\!\left|h_{m,n,H}\right|^2\!\left(1\!+\!\gamma_{m,n}^{1}\right)}{\sum N_m}}.
\end{align}

\indent Similarly, by setting $\frac{\partial L}{\partial B_{m}^{T}}=0$, we have
\begin{align}
&-\lambda_B+X_{B_{m}^{T}}+Y_{B_{m}^{T}}\times\frac{1}{\sqrt{B_{m}^{T}}}=0,\label{2a}
\end{align}
where $Y_{B_{m}^{T}}=\sum\limits_{n=1}^{N_m} y_{m,n}^{2}\sqrt{\frac{1}{N_m}P_{m,n}^{T}\left|h_{m,n,T_m}\right|^2\left(1+\gamma_{m,n}^{2}\right)}$ and $X_{B_{m}^{T}}=\frac{1}{N_m}\sum\limits_{n=1}^{N_m} \log_2(1+\gamma_{m,n}^{2})-\gamma_{m,n}^{2}$.\\
\indent Combining (\ref{1a}), (\ref{2a}) and (\ref{pro1e}), we have
\begin{equation}
(\frac{Y_{B_{H}}}{\lambda_B^{\ast}-X_{B_{H}}})^2+\sum_{m=1}^M (\frac{Y_{B_{m}^{T}}}{\lambda_B^{\ast}-X_{B_{m}^{T}}})^2-B_C=0.\label{c1}
\end{equation}
Bases on (\ref{c1}), $\lambda_B^{\ast}$ can be obtained by the line search method. $B_{H}^{\ast}$ and ${B_{m}^{T}}^{\ast}$ can be obtained by solving (\ref{1a}) and (\ref{2a}), i.e.,
\begin{equation}\label{B_Tm_star}
B_{H}^{\ast}=(\frac{Y_{B_{H}}}{\lambda_B^{\ast}-X_{B_{H}}})^2,\quad{B_{m}^{T}}^{\ast}=(\frac{Y_{B_{m}^{T}}}{\lambda_B^{\ast}-X_{B_{m}^{T}}})^2.
\end{equation}
\indent After determining the optimal bandwidth allocation scheme, we determine the optimal power allocation scheme. Similarly, setting $\frac{\partial L}{\partial P_{m,n}^{T}}=0$, the optimal ${P_{m,n}^{T}}^{\ast}$ can be given by
\begin{equation}
{P_{m,n}^{T}}^{\ast}=(\frac{G_{m,n}^{T}}{\lambda_{T,m}^{\ast}-F_{m,n}^{T}})^2,\label{PUT_star}
\end{equation}
where $G_{m,n}^{T}=y_{m,n}^{2}\sqrt{\frac{B_{m}^{T}}{N_m}\left|h_{m,n,T_m}\right|^2\left(1+\gamma_{m,n}^{2}\right)}$ and
\begin{equation}\label{FTmn}
\begin{split}
F_{m,n}^{T}\!&=\!\sum_{\substack{n^{\prime}\in N_m\\n^{\prime}\neq{n}}}\!(-{y_{m,n^{\prime}}^{1}}^2\!\times\!\frac{1}{J}|h_{m,n',T_m}|^2)\!-\!{y_{m,n}^{2}}^2|h_{m,n,T_m}|^2\\
&\quad \!+\!\sum_{m'\neq m}\sum_{n'=1}^{N_{m'}}\!\left(\left(\!-{y_{m^{\prime},n^{\prime}}^{1}}^2\!-\!y_{2,m',n'}^2\right)\!\times\!\frac{1}{J}|h_{m^{\prime},n^{\prime},T_m}|^2\right).
\end{split}
\end{equation}

\begin{algorithm}[t]
	\label{alg3}
	\caption{Bandwidth and Power Allocation}
	\KwIn{$P_{H}$, $P_{H,Ka}$, $\max\{C_{TH}+C_{TS}\}$, $M_H$, $M_S$;}
	\textbf{Initialization}: $B_{m}^{T}$, $B_{H}$, $P_{m,n}^{H}$, $P_{m,n}^{T}$;\\
	\Repeat{Convergence}{
	\textbf{1)} Compute ${\gamma_{m,n}^{1}}^{\ast}$ and ${\gamma_{m,n}^{2}}^{\ast}$ by (\ref{gamma_star});\\
	\textbf{2)} Compute ${y_{m,n}^{1}}^{\ast}$ and ${y_{m,n}^{2}}^{\ast}$ by (\ref{y_star});\\
	\textbf{3)} Search for $\lambda_B^{\ast}$ by (\ref{c1}). Same for $\lambda_{T,m}^{\ast}$, $\lambda_H^{\ast}$; \\
	\textbf{4)} Compute ${B_{m}^{T}}^{\ast}$, $B_{H}^{\ast}$, ${P_{m,n}^{H}}^{\ast}$, ${P_{m,n}^{T}}^{\ast}$ by (\ref{B_Tm_star}), (\ref{PHC_star}), (\ref{PUT_star}), respectively;\\
	\textbf{5)} Compute $R_H$, $R_T$ by (\ref{RH}), (\ref{RT});
	}
    \KwOut{$B_{m}^{T}$, $B_{H}$, $P_{m,n}^{H}$, $P_{m,n}^{T}$;}
\end{algorithm}
\noindent $\lambda_{T,m}^{\ast}$ can be obtained by the line search method based on $\sum\limits_{n} (\frac{G_{m,n}^{T}}{\lambda_{T,m}^{\ast}-F_{m,n}^{T}})^2-P_{T}=0$.\\
\indent Setting $\frac{\partial L}{\partial P_{m,n}^{H}}=0$, the optimal ${P_{m,n}^{H}}^{\ast}$ can be given by
\begin{equation}
{P_{m,n}^{H}}^{\ast}=(\frac{G_{m,n}^{H}}{\lambda_H^{\ast}-F_{m,n}^{H}})^2,\label{PHC_star}
\end{equation}
where $G_{m,n}^{H}=y_{m,n}^{1}\sqrt{\frac{B_{H}}{\sum_{m=1}^M N_m}\left|h_{m,n,H}\right|^2\left(1+\gamma_{m,n}^{1}\right)}$
and
\begin{equation}
F_{m,n}^{H}\!=\!-{y_{m,n}^{1}}^2|h_{m,n,H}|^2\!+\!\frac 1J\!\sum_{m^{\prime},n^{\prime}}\!(-{y_{m,n}^{2}}^2)\vert h_{m^{\prime},n^{\prime},H}\vert^2.
\end{equation}

\noindent $\lambda_H^{\ast}$ can be obtained by the line search method based on $\sum\limits_{m,n} (\frac{G_{m,n}^{H}}{\lambda_H^{\ast}-F_{m,n}^{H}})^2-P_H+P_{H,Ka}\times M_H=0$.\\
\indent The optimization process consists of multiple iterations. In each iteration, Lagrange multipliers are updated first, $B_{m}^{T}$, $B_{H}$, $P_{m,n}^{H}$, $P_{m,n}^{T}$ are then updated accordingly. The iterative sum-rate optimization algorithm is summarized in Algorithm \ref{alg3}.

\section{Simulation Results}\label{SR}
In this section, we evaluate the performance of the proposed sum-rate maximization algorithm. Major simulation parameters are specified based on the existing works \cite{b7} and 3GPP specifications \cite{b8} as given in Table \ref{parameter}.\\
\begin{table}[!htbp]
\footnotesize
\renewcommand\arraystretch{1.3}
  \centering
  \caption{Simulation Parameters}
  \label{parameter}
  \vspace{-0.6cm}
  \begin{tabular}{c|c}
     \hline
     \hline
     \textbf{Parameters} & \textbf{Values} \\
     \hline
     \hline
     Bandwidth for C-band communications $B_C$(MHz) & $20$ \\
     \hline
     Noise density for C-band communications $\sigma_c^2$(dBm/Hz) & $-174$\\
     \hline
     Transmit power of each UT $P_{T}$ (W) & $20$ \\
     \hline
     Power of the HAP $P_{H}$ (W) & $60$ \\
     \hline
     Power of the satellite $P_{SAT}$ (W) & $60$ \\
     \hline
     Number of orthogonal channels on C-band $J$ & $1000$\\
     \hline
     Antenna amplify gain of HAP $G_H$ (dB) & $20$ \\
     \hline
     Antenna amplify gain of LEO satellites $G_S$ (dB) & $27$ \\
     \hline
     Backhaul capacity of HAP over C-band $C_H$ (Mbps) & $17.5$\\
     \hline
     Altitude of HAP (km) & $20$\\
     \hline
     Altitude of LEO satellites (km) & $200$\\
     \hline
     Bandwidth for Ka-band communications $B_{Ka}$ (MHz) & $800$ \\
     \hline
     Noise density for Ka-band communications $\sigma^2$ (dBm/Hz) & $-203$\\
     \hline
     \hline
   \end{tabular}
   \vspace{-0.3cm}

\end{table}

\begin{figure}
  \centering
  \includegraphics[width=3.5in]{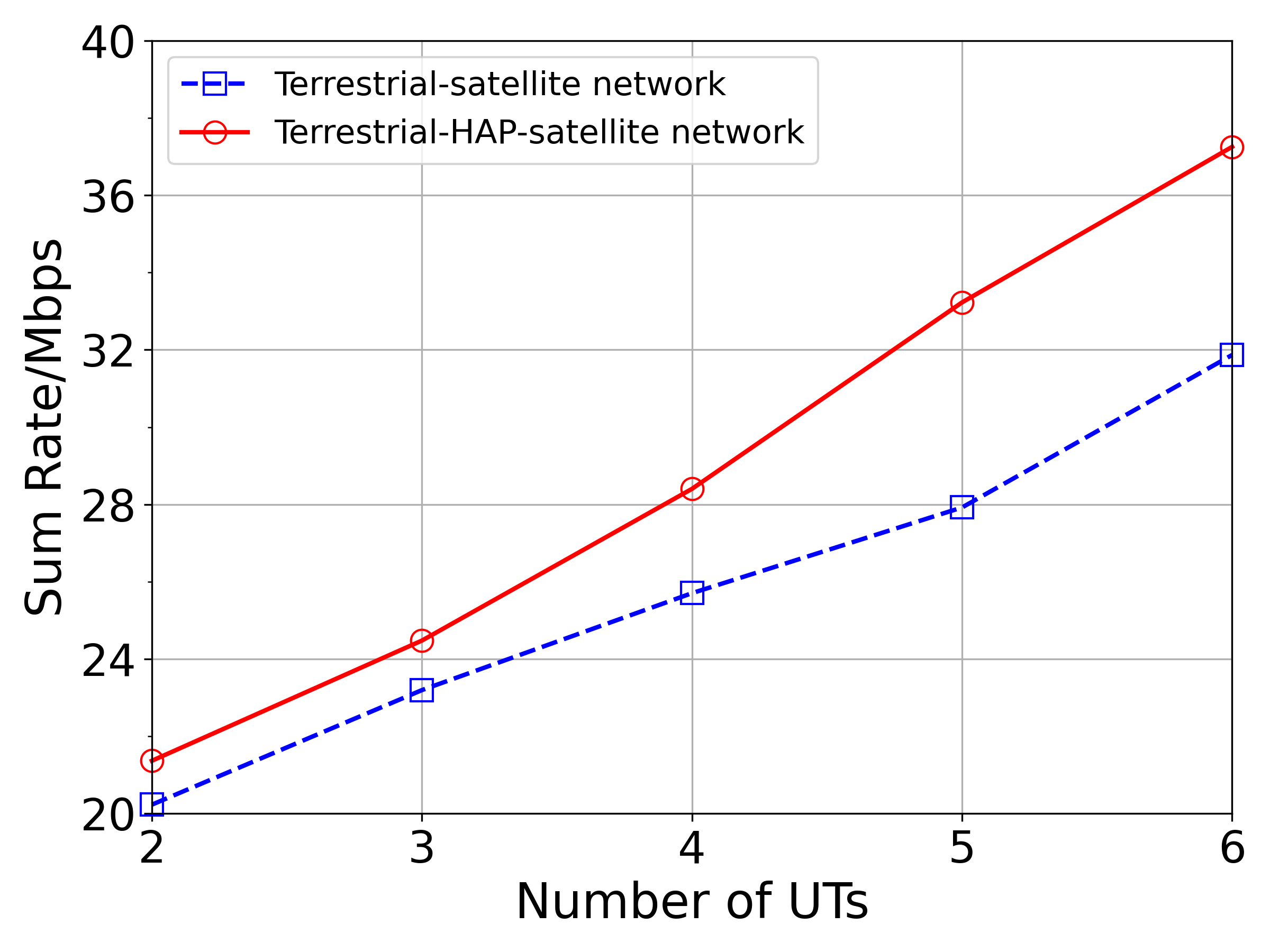}\\
  \caption{Sum rate versus the number of UTs}
  \label{Fig1}
\end{figure}

\indent Fig. \ref{Fig1} shows the sum rate versus the number of UTs. It can be seen that the sum rate increases with the number of UTs, i.e., $M$, since the power gain increases with $M$. We also observe that the three-layer integrated satellite-HAP-terrestrial network performs better than the two-layer integrated terrestrial-satellite network. The main reason is that the three-layer network where the HAP supports both C-band and Ka-band communications can provide more access methods for users.\\

\begin{figure}
  \centering
  \includegraphics[width=3.5in]{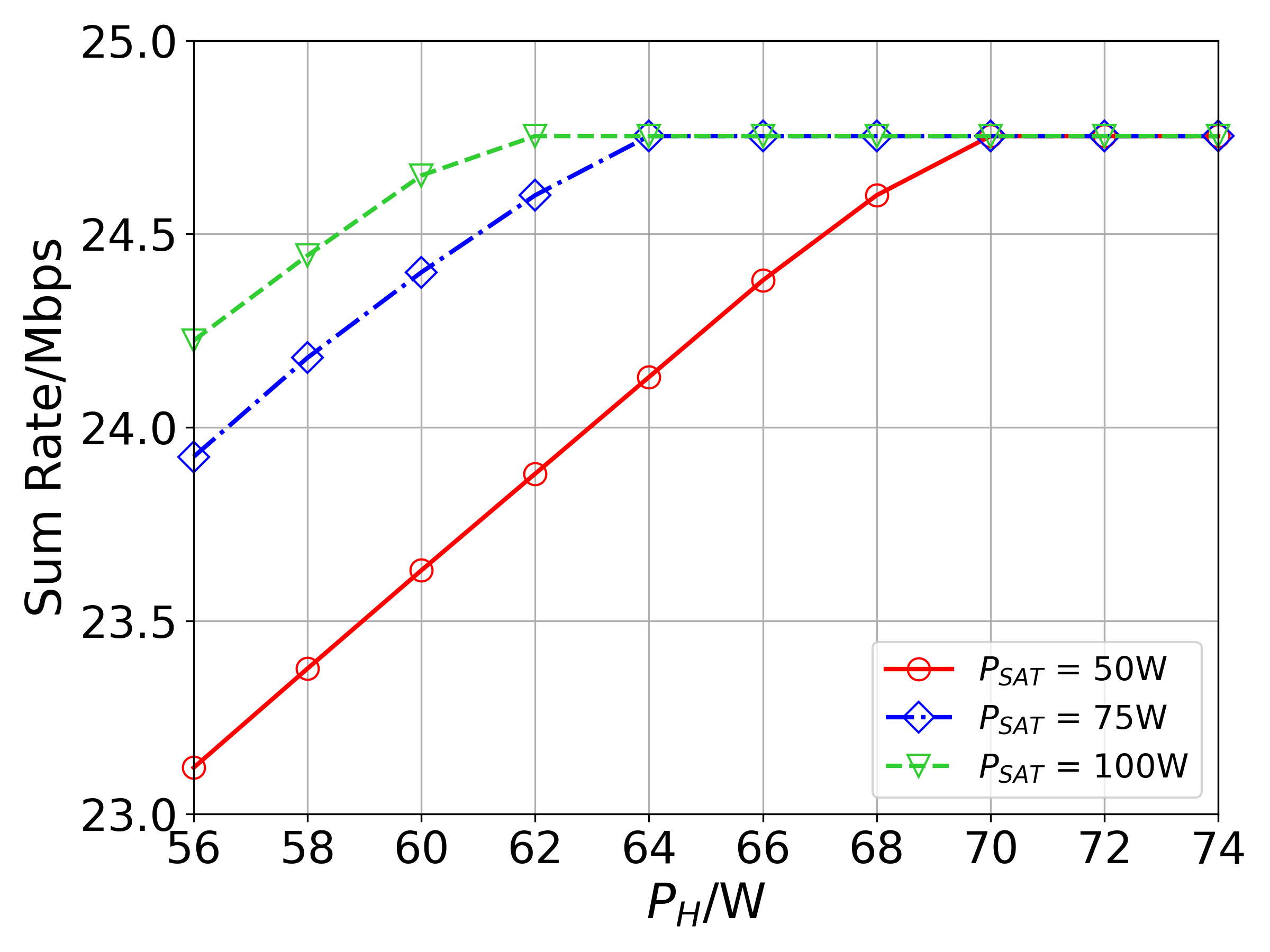}\\
  \caption{Sum rate versus the power of HAP (The number of UTs $M=3$}
  \label{Fig3}
\end{figure}

\indent Fig. \ref{Fig3} illustrates the sum rate versus the power of HAP, i.e., $P_H$. The sum rate first increases and then saturates as $P_H$ grows. The main reason for saturation is that the sum rate is upper bounded by the backhaul capacity of the HAP over C-band and the total transmits power provided by all UTs. It can also be seen that the sum rate grows with the power of the satellite. However, the maximum value of the sum rate is irrelevant to the power of the satellite, since when $P_{H}$ is large enough, all UTs will connect with HAP for Ka-band backhaul.\\

\begin{figure}
  \centering
  \includegraphics[width=3.5in]{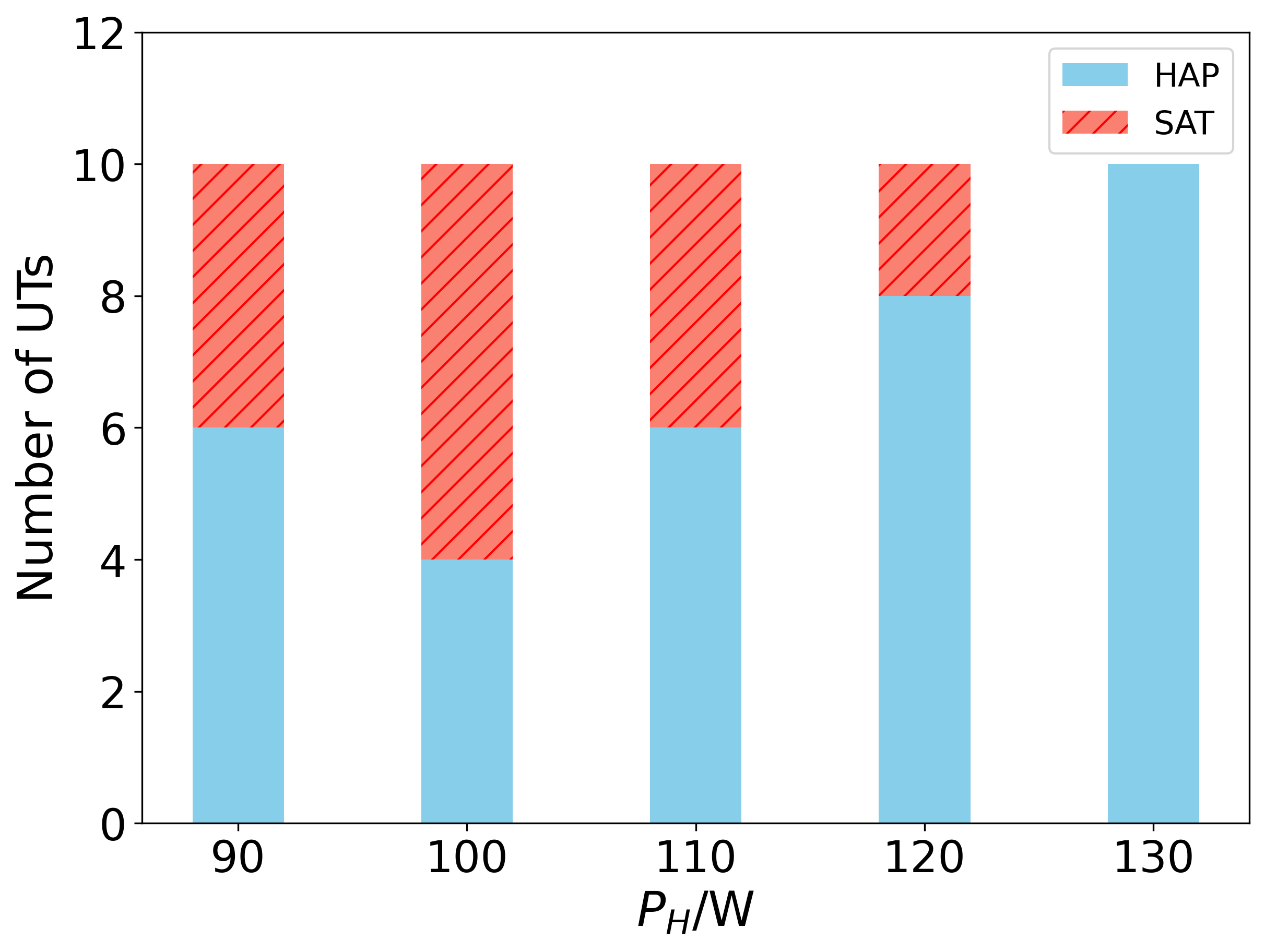}\\
  \caption{Distribution of UTs' connections for Ka-band backhaul versus the power of HAP}
  \label{Fig2}
\end{figure}

\indent Fig. \ref{Fig2} shows the distribution of UTs' connections for Ka-band backhaul versus the total power of HAP, i.e., $P_H$. It can be seen that the number of UTs connecting with the HAP first decreases and then increases as $P_H$ grows. The main reason is that the HAP allocates more power for direct communications to users over C-band as $P_H$ grows such that the power allocated to Ka-band, i.e., $P_{H,Ka}$ decreases as $P_H$ grows. The number of UTs connecting with the HAP for Ka-band backhaul then decreases. However, when $P_H$ is large enough, $P_{H,Ka}$ also increases as $P_H$ grows, thereby increasing the number of UTs connected with the HAP for Ka-band backhaul.

\section{Conclusions}\label{Con}
In this paper, we have investigated resource allocation in an integrated three-layer satellite-HAP-terrestrial network to exploit wireless communication enhancement brought by dual-band connectivity of the HAP. A sum-rate maximization problem has been formulated. To solve the problem efficiently, we decompose the problem into two subproblems, i.e., the backhaul capacity maximization problem and the resource allocation problem. An iterative optimization algorithm has been developed to solve these two subproblems iteratively. Specifically, in each iteration of the proposed algorithm, the backhaul capacity of UTs has been maximized first through dynamic programming. Following that, the resource allocation scheme has been optimized via fractional programming. The closed-form optimal solution for bandwidth allocation and power allocation in each iteration is also derived. Simulation results show that the integration of the HAP into the satellite-terrestrial network brings extra gain due to the dual-band connectivity of the HAP. Besides, the backhaul selection of the UT is relevant to the transmit power of the HAP.

\footnotesize


\begin{thebibliography}{40}

\bibitem{b01}
R. Deng \emph{et al.}, "Ultra-dense LEO satellite constellations: How many LEO satellites do we need?" \emph{IEEE Trans. Wireless Commun.}, Early access, doi: 10.1109/TWC.2021.3062658.

\bibitem{SHB-2019}
S. Zhang, H. Zhang, B. Di, and L. Song, ``Cellular UAV-to-X communications: Design and optimization for multi-UAV networks," \emph{IEEE Trans. Wireless Commun.}, vol. 18, no. 2, pp. 1346-1359, Feb. 2019.


\bibitem{b0}
Federal Communications Commissions, \emph{SpaceX non-geostationary satellite system (Attachment A)}, 2016.



\bibitem{b1}
Alejandro Aragón-Zavala \emph{et al.}, ``Overview on HAPS,'' in \emph{High-Altitude Platforms for Wireless Communications}, 1st ed. New York:Wiley, ch.2, sec.2, 2008, pp.5-35.

\bibitem{b2}
A. Mohammed \emph{et al.}, ``The role of high-altitude platforms (HAPs) in the global wireless connectivity,'' \emph{Proc. IEEE}, vol. 99, no. 11, pp. 1939-1953, Nov. 2011.

\bibitem{b3}
D. Yuniarti, ``Regulatory challenges of broadband communication services from High Altitude Platforms (HAPs),'' in \emph{Int. Conf. Inf. Commun. Technol.}, Yogyakarta, Indonesia, 2018, pp. 919-922.

\bibitem{b31}
B. Di \emph{et al.}, ``Ultra-dense LEO: Integrating terrestrial-satellite networks into 5G and beyond for data offloading,'' \emph{IEEE Trans. Wireless Commun.}, vol. 18, no. 1, pp. 47-62, Jan. 2019.

\bibitem{b4}
D. Xu \emph{et al.}, ``Coverage ratio optimization for HAP communications,'' in \emph{IEEE Int. Symp. Person Indoor Mobile Radio Commun.}, Montreal, QC, Canada, 2017, pp. 1-5.

\bibitem{b5}
A. Ibrahim and A. S. Alfa, ``Using Lagrangian relaxation for radio resource allocation in high altitude platforms,'' \emph{IEEE Trans. Wireless Commun.}, vol. 14, no. 10, pp. 5823-5835, Oct. 2015.

\bibitem{b6}
Y. Li \emph{et al.}, ``An extensible multi-layer architecture model based on LEO-MSS and performance analysis,'' in \emph{IEEE Veh. Technol. Conf.}, Honolulu, HI, USA, 2019, pp. 1-6.

\bibitem{b9}
K. Shen and W. Yu, ``Load and interference aware joint cell association and user scheduling in uplink cellular networks'' in \emph{IEEE Veh. Technol. Conf.}, Edinburgh, UK, 2016, pp. 1-5.

\bibitem{b7}
R. Deng \emph{et al.}, ``Ultra-dense LEO satellite offloading for terrestrial networks: How much to pay the satellite operator?'' \emph{IEEE Trans. Wireless Commun.}, vol. 19, no. 10, pp. 6240-6254, Oct. 2020.

\bibitem{b8}
3GPP TR 38.811 (V0.3.0), Study on new radio (NR) to support non terrestrial network (Release 15), Dec. 2017.
\end{thebibliography}
\end{document}